\documentstyle[twocolumn,prl,aps]{revtex}
\begin{document}
\twocolumn[\hsize\textwidth\columnwidth\hsize\csname
  @twocolumnfalse\endcsname
\begin{center}
{\sc To be published in Europhysics Letters}
\end{center}
\title{Spin Polarizations at and about the Lowest Filled Landau
Level}
\author{Tapash Chakraborty}
\address{Institute of Mathematical Sciences, Taramani, Madras
600 113, India}
\author{P. Pietil\"ainen}
\address{Theoretical Physics, University of Oulu,
Linnanmaa, FIN-90570 Oulu, Finland}
\author{R. Shankar}
\address{Institute of Mathematical Sciences, Taramani, Madras
600 113, India}
\date{6 January 1997}
\maketitle

\begin{abstract}
The spin polarization versus temperature at or near a fully
filled lowest Landau level is explored for finite-size systems in
a periodic rectangular geometry. Our results at $\nu=1$ which
also include the finite-thickness correction are in good
agreement with the experimental results. We also find that the
interacting electron system results are in complete agreement
with the results of the sigma model, i.e., skyrmions on a torus
have a topological charge of $Q\ge2$ and the $Q=1$ solution is
like a single spin-flip excitation. Our results therefore
provide direct evidence for the skyrmionic nature of the
excitations at this filling factor.
\end{abstract}
  \vskip 2pc ] 
\narrowtext
At the Landau level filling factor $\nu=1$ $(\nu=N_e/N_s$
where $N_e$ is the electron number and $N_s=AeB/hc=A/2\pi\ell_0
^2$ is the Landau level degeneracy and $\ell_0$ is the
magnetic length) the ground state is fully spin polarized with
total spin $S=N_e/2$ \cite{zhang}. In recent Knight-shift spin
polarization measurements \cite{barrett}, a precipitous fall in
the spin polarization was observed when either one moves slightly
away from $\nu=1$ or the temperature is increased at $\nu=1$.
This effect has also been observed in subsequent experiments with
tilted magnetic field as well as optical absorption studies
\cite{others}. Theoretically, such a result is explained as due
to the fact that the low-energy charged excitations are spin
textures (skyrmions) \cite{sondhi,fertig} instead of the single
spin-flip excitations \cite{zhang} (the latter excitations are,
of course, possible only for large values of the Zeeman energy,
or large values of the $g$-factor). In fact, finite-size
system calculations in a spherical geometry indicated that,
when one adds or removes a flux quantum at $\nu=1$ the total
spin changes to $S=0$ \cite{sondhi}.
 
Theoretical studies at $\nu=1$ indicated that \cite{sondhi}
for large values of $g$ the excitations are of single-particle
type, i.e., they carry charge $\pm e$ and spin $S_z=\frac12$
and they have the size of magnetic length $\ell_0$. As $g$ is
decreased, the excitations still carry charge $\pm e$, but they
cover an extended region and have a nontrivial spin order: at the
boundary of the system the local spin takes the value of the
ground state and reversed at the center of the skyrmion. Along
any radius, the spin gradually twists between these two limits.
The size of the skyrmion is determined by the competition between
the interaction energy and the Zeeman energy. The former favors a
large size in order to have uniform charge density, while the
latter, with increasing strength, tends to reduce the size.
 
Earlier theoretical studies of spin polarization versus
temperature at $\nu=1$ involved a continuum quantum field theory
of a ferromagnet as a model and its properties at finite
temperatures \cite{field}. The other work was based on a
many-body perturbation theory \cite{rpa}. A qualitative agreement
between the calculated temperature dependence of the spin 
polarization from these theories and the observed results was
achieved. 
 
We have employed finite-size systems in a periodic rectangular
geometry to study the spin polarization $\langle S_z\rangle$
as a function of temperature at and about $\nu=1$. Here
$$\langle S_z(T)\rangle\equiv\frac1Z\langle0|S_z|0\rangle+
\sum\frac1Z\,e^{-\varepsilon_j/kT}\langle j|S_z|j\rangle,$$
where $|0\rangle$ is the ground state, $Z=\sum_j\ e^{-\beta
\varepsilon_j}$ is the partition function and the summation is
over all excited states $|j\rangle$ with energy $\varepsilon_j$.
The ground state and the excited states are calculated from the
exact diagonalization of a few-electron system Hamiltonian in a
periodic rectangular geometry \cite{tapash}.
 
The results for $\langle S_z(T)\rangle$ versus $T$ (in units of
$e^2/\epsilon\ell_0$ where $\epsilon$ is the background
dielectric constant) in an eight-electron system at $\nu=1$ is
plotted in Fig. 1(a) where the magnetic field is held fixed 
$(B=10$T) but the $g$-factor is varied $(0.1-0.5)$. In recent
experiments \cite{pressure} it was shown that applying
hydrostatic pressure on the electron system, one can vary the 
$g$-factor at a given magnetic field. The experimental results 
by Barrett et al. \cite{barrett} are also plotted in Fig. 1 for 
comparison. The observed
data show a much sharper drop with increasing temperature than
the theoretical results obtained here. As mentioned above, the
size of spin-excitations is dictated by the competition between
the Zeeman energy (which is controlled here by the $g$-factor)
and the interaction energy. Interestingly, the interaction
potential can also be modified by including the finite-thickness
correction to the electron-electron interaction, in the 
calculation \cite{thickness,book}. This is
shown in Fig. 1(b) where we present the results for a 
finite-thickness correction parameter $\beta=0.5$ as an example
\cite{thickness,book}. The agreement with the experimental
results now improves noticeably. The changes in our results are
most pronounced for large $g$, which is a direct evidence of the
competition between the Coulomb and Zeeman energies mentioned
above. Our results are also in good agreement with recent
magnetoabsorption spectroscopy results \cite{bennett} (Fig. 1).
 
The skyrmion description of quasiparticle (hole) excitations near
$\nu=1$ \cite{sondhi,lee} assumes that the low-energy, 
long-wavelength effective Hamiltonian of the system is given by
\begin{eqnarray}
{\cal H}=&&\int d{\bf x}\ \gamma\ \partial_i{\bf n}(x) \cdot
\partial_i{\bf n}(x)\nonumber \\
+\frac12 &&\int d{\bf x}\int d{\bf y}\ q(x)\ 
U(x,y)\ q(y).
\label{ham}
\end{eqnarray}
Here $\bf{\hat n}(x)$ is a unit vector field representing the 
local spin polarization, $U(x,y)$ is the inter-electron 
interaction potential and $\gamma$ is a constant which is related
to the spin stiffness. In our following arguments we shall 
consider the $g=0$ case. The deviation of the electron charge 
density from its $\nu=1$ value is given in terms of the spin 
density
\begin{equation}
q(x)=\frac1{4\pi}\epsilon_{ij}\, {\bf n}(x)\cdot\left(\partial_i
{\bf n}(x)\times\partial_j {\bf n}(x)\right),
\label{topcharge}
\end{equation}
where the right hand side is the topological charge density. It 
is to be noted that $Q=\int_x q(x)$ is always an integer which is
equal to the number of times $\bf{\hat n}(x)$ wraps around the
sphere as $\bf x$  varies over all space. 
 
In Eq. (\ref{ham}), both the terms are small if ${\bf n}(x)$ is 
slowly varying i.e., if its derivatives are small. We therefore
expect that {\it smoother} the configuration, smaller its energy 
would be. The smoothest spin configuration that one can imagine 
on any closed surface (e.g., a sphere or a torus) is the {\it 
hedgehog} configuration. That is the spin configuration where the
spin is always normal to the surface. As we go over all space of 
a sphere the spin density vector $\bf n$ covers the sphere 
exactly once. The charge of this state is therefore $Q=1$ and 
the total spin is zero. This effect has been observed numerically
in the interacting electron systems \cite{sondhi}.
 
Let us consider the electron system on a plane with periodic boundary 
conditions, neglect the interaction term initially and consider 
the energy functional of the sigma model [the first term of Eq. 
(\ref{ham})]. All the multisoliton solutions of the sigma model 
are exactly known \cite{polya}. In terms of the variables 
$w=\cot({\theta\over 2})e^{i\phi}$ (which correspond to the 
stereographic projection of the sphere onto the complex plane) 
the solutions are analytic and antianalytic functions. They 
saturate the Bogomol'nyi bound for the energy which is given by
\begin{equation}
E\ge8\pi\gamma|Q|.
\label{bb}
\end{equation}
For periodic boundary conditions the solutions are doubly
periodic analytic functions, namely elliptic functions. A basic
property of elliptic functions is that the sum of its residues at
its poles in the fundamental domain is zero \cite{witt}. The
charge $Q$ of the configuration is therefore at least two.
The winding number is calculated as \cite{polya} $Q=\sum_i n_i$,
where $i$ runs over all the poles and $n_i$ is the order of the
$i$-th pole. Now, if the sum of the residues is to vanish, then
the elliptic function must have at least two simple poles
or a second-order pole. Therefore we must have $Q \ge 2$ for the 
solutions in this geometry. Pictorially, this would mean that the
hedgehog on a torus has $Q=0$ -- the sphere is covered twice as 
we go over a torus, but it is covered in the opposite sense such 
that the total winding number is zero. In other words, the 
hedgehog on a torus can be viewed as a {\it skyrmion-antiskyrmion
pair}. But if we reflect all the spins in the inner half of the 
torus about the $xz$ (or $yz$) plane, then the sign of the 
topological charge density flips in the inner half and we get  
$Q=2$. The total spin of both the above configurations is zero 
by symmetry.
 
In what follows, we show that the $Q=1$ configurations that 
saturate the bound in Eq. (\ref{bb}) are singular and pointlike.
Consider the configuration $w(z,\bar z)=A~{\rm sn}(z)~e^{i\Omega
(z,\bar z)}$, where ${\rm sn}(z)$ is one of
the Jacobian elliptic functions with periods $4K,2iK'$
\cite{witt}, and $A$ is a constant. If we choose $2K=L_x$ and
$2K'=L_y$, where $L_x$ and $L_y$ are the lengths of the sides of
the rectangle, then ${\rm sn}(z)$ is periodic in the $y$
direction and antiperiodic in the $x$ direction. If
$\Omega(z+L_x,\bar z+L_x)=\Omega(z,\bar z)+\pi$ and it is
periodic in the $y$ direction then $w(z,\bar z)$ is periodic. It
is easily verified that if the derivatives of $\Omega$ are
periodic then $w(z,\bar z)$ has $Q=1$. The energy of $w$ is
$$E[w]=8\pi\gamma+4\gamma\int d{\bf x}~{(A|{\rm sn}(z)|)^2 \over
\left(1+(A|{\rm sn}(z)|)^2\right)^2}~\left(\partial_i\Omega
\right)^2. $$
In the limit $A\rightarrow0$, the first factor of the integrand
is sharply peaked about the pole of ${\rm sn}(z)$. Therefore, if 
the derivatives of $\Omega$ vanish near that point then the bound
(\ref{bb}) is saturated in the above limit. The $Q=1$
configuration that minimizes the energy is therefore singular and
pointlike. Inclusion of the Coulomb interaction is likely to make
it regular but it will still be localized in a small region near 
the pole. We therefore conclude that the $Q=1$ excitation in the 
rectangular periodic geometry is more like a single spin-flip 
excitation and not a spin texture. We should point out here that 
the above arguments apply only to the $g=0$ case where the 
solutions for $w$ are analytic (and antianalytic) and therefore 
sensitive to boundary conditions.
 
We now argue that in the presence of Coulomb repulsion the $Q=2$ 
solutions will have {\it zero spin}. We take as an ansatz, 
$w(z)~=~A~{\rm sn}(z)$, where $L_x=4K$ and $L_y=2K'$.
The Jacobian elliptic functions have well-seperated poles and
zeros. Since the poles and zeros correspond to north and south
poles respectively these correspond to slowly varying spin
configurations. Further, if we choose $A=\sqrt k$, where $k$
is the modulus \cite{witt}, then using the property,
${\rm sn}(z+iK')=\left[k\ {\rm sn}(z)\right]^{-1}$, we have
$w\left(z+i{L_y/2}\right)=\left[w(z)\right]^{-1}$, and as a
result, $q\left(z+i{L_y/2}\right)=q(z)$. This choice of $A$
would give the lowest value of the coulomb energy since
the charge is spread out symmetrically. Since
$\cos\theta=(w\bar w-1)/(w\bar w+1)$, the transformation property
of $w(z)$ mentioned above can be used to show that the $z$
component of the total spin vanishes. The $x$ and $y$ components
of the total spin can be shown to be zero using the property that
${\rm sn}(z+2K)=-{\rm sn}(z)$. The periodic rectangular geometry 
therefore provides a unique test for the geometrical and 
topological aspects of the sigma model scenario. It predicts that
the spin of the $Q=1$ excitations should not deviate very much 
from the ground state spin whereas the spin of the $Q=2$ 
excitation should drop to zero.
 
Our exact diagonalization studies provide strong support for 
these predictions. In Fig. 2, we present the results for 
$\langle S_z(T)\rangle$ at $\nu=1$ with one flux
quantum added ($\nu=8/9$) or removed ($\nu=8/7$). Obviously, the
spin polarization does not drop to zero, in contrast to what one
expects in a spherical geometry, but has the nature of a single
spin-flip excitation as anticipated above. The rapid drop in spin
polarization takes place only when we add or remove {\it two}
flux quanta from the system i.e., at $\nu=8/10$ or at $\nu=8/6$.
Indeed, for both fractions $\langle S_z(T\sim0)\rangle=0$ when
$g$ is small (Fig. 3). It should be pointed out however, that
both $\nu=\frac45$ and $\nu=\frac43$ are genuine fractional
quantum Hall states which were observed experimentally
\cite{book,fqhe}. Additionally, for the latter fraction, the fact
that it has $S=0$ was already established theoretically as well
as experimentally at $T=0$ \cite{book}. It is quite difficult to
envision these many-body states as two-skyrmion excitations of
$\nu=1$. Although desirable, it is a formidable task to
diagonalize even larger systems. Therefore we opted for a
six-electron system instead. Here, for $\nu=6/8$ or $\nu=6/4$ we
do not expect, a priori, that the ground state has $S=0$.
However, as shown in Fig. 4, both these states have $S=0$ at
$T=0$ and they also show other expected behavior of a
spin-singlet state \cite{tapash,vonK}. Observation of such a 
state at these fractions therefore lends support to our 
prediction above that toroidal geometry supports at least two 
skyrmions as lowest-energy spin excitations at $\nu=1$.
 
In conclusion, at $\nu=1$ our finite-size system results are in
good agreement with the experimental results. Finite-thickness 
correction of the interaction is found to improve the agreement,
thereby providing strong support for the skyrmionic picture of
excitations near $\nu=1$. We have also demonstrated that the
results of the interacting electron system are in complete
agreement with the sigma model predictions.

One of us (TC) would like to thankfully acknowledge the
hospitality of Scuola Normale Superiore, Pisa, and International
Center for Theoretical Physics, Trieste, where part of the work
was done. He also thanks Sean Barrett and Bennett Goldberg for
communicating their experimental results.

\newpage

\begin{figure}
\caption{Electron spin polarization $\langle S_z (T)\rangle$ as
a function of temperature $T$ at $\nu=1$ for a 8-electron system
without (a) and with (b) finite-thickness correction included.
Experimental results of \protect\cite{barrett} $(\Box)$ and
\protect\cite{bennett} $(\circ,\bullet)$ are also given for
comparison.
}
\label{fig1}
\end{figure}
\begin{figure}
\caption{Electron spin polarization versus $T$ at $\nu=1$ for a
8-electron system with one flux quantum added or removed.
}
\label{fig2}
\end{figure}
\begin{figure}
\caption{Electron spin polarization versus $T$ at $\nu=1$ for a
8-electron system with two flux quanta added or removed.
}
\label{fig3}
\end{figure}
\begin{figure}
\caption{Same as in Fig. 3 but for a 6-electron system.
}
\label{fig4}
\end{figure}
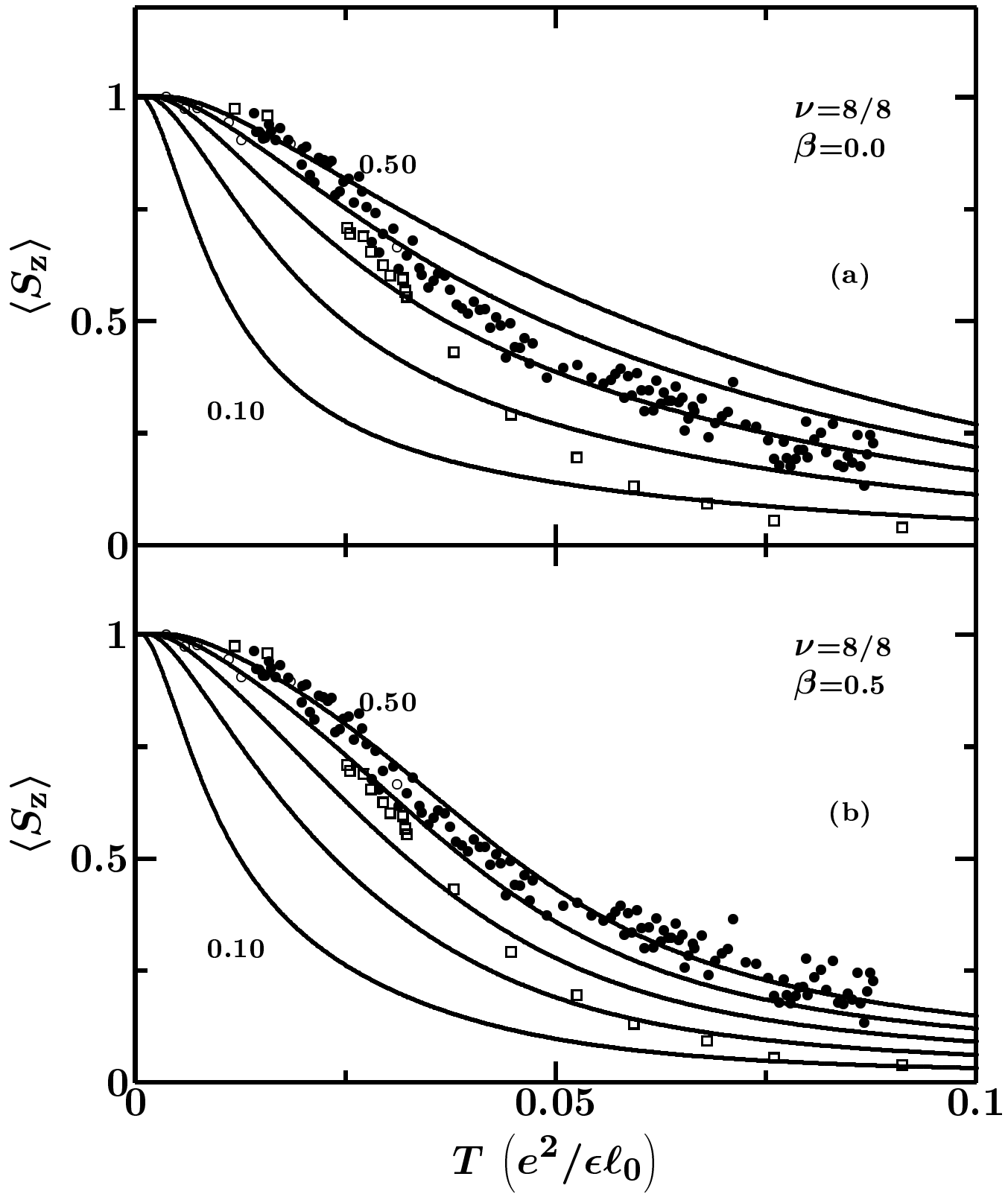
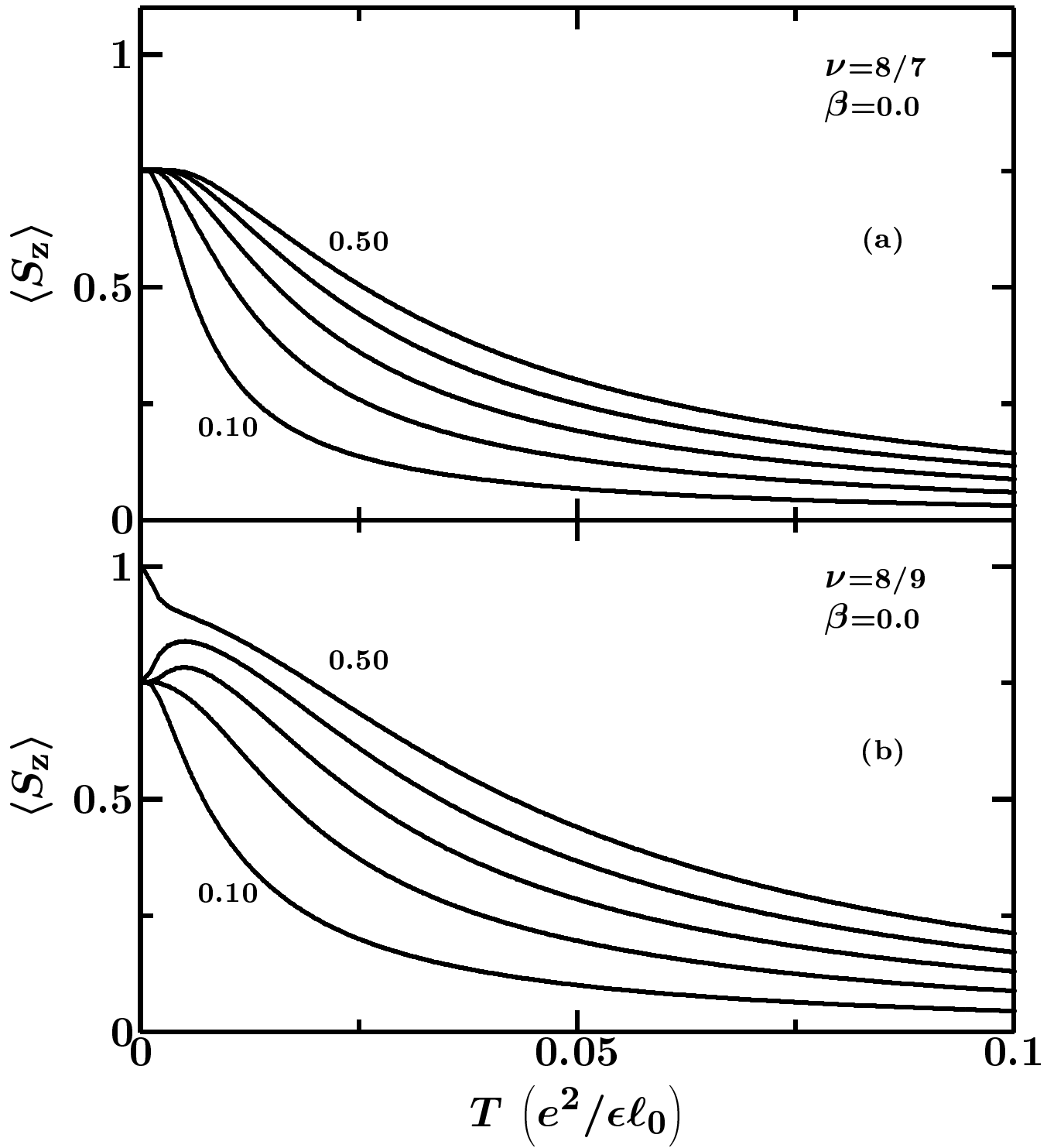
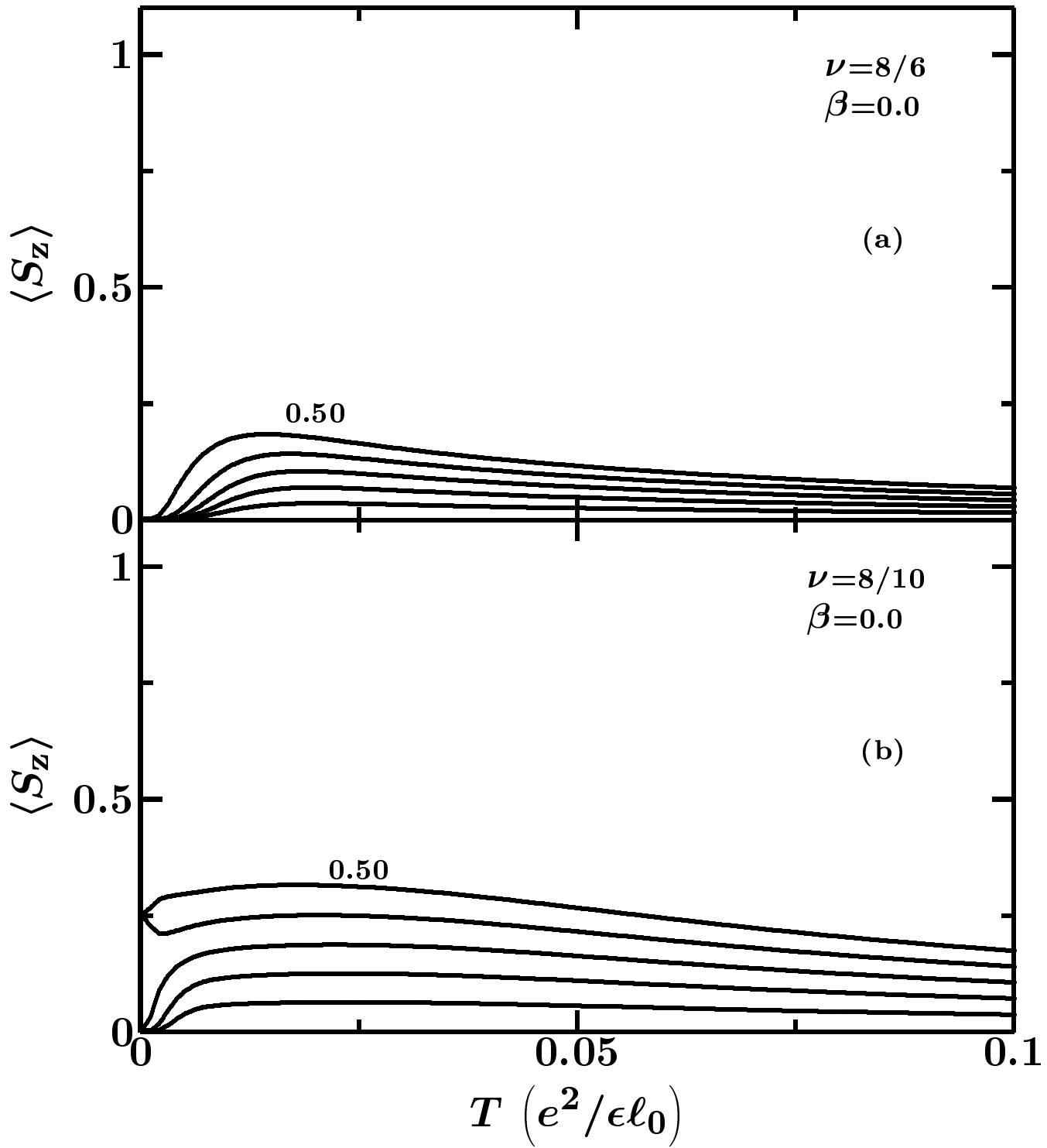
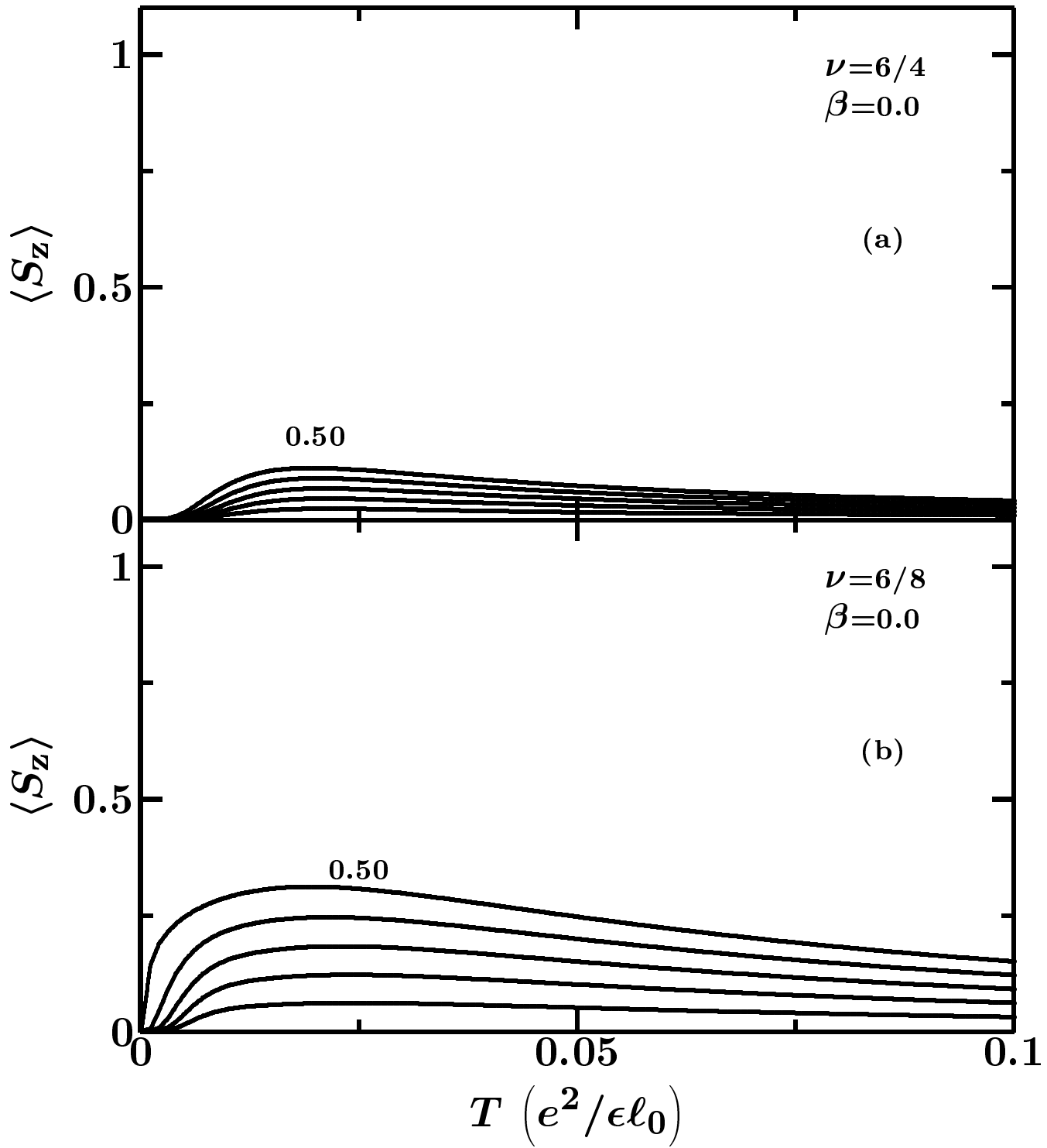
\end{document}